\documentclass[
 reprint,
 amsmath,amssymb,
 aps,
 showkeys
]{revtex4-2}

\usepackage{graphicx}
\usepackage{dcolumn}
\usepackage{bm}
\usepackage{times}
\usepackage[colorlinks]{hyperref}

\begin{document}

\preprint{APS/123-QED}

\title{Resonant attenuation of surface acoustic waves by a weakly bonded layer}

\author{Martin Robin}
\email{martin.robin@lspm.cnrs.fr, now affiliated to Laboratoire des Sciences des Procédés et des Matériaux (LSPM), CNRS, Université Sorbonne Paris Nord, 93430, Villetaneuse, France}
\author{Thomas Dehoux}
\author{Maroun Abi Ghanem}
 \email{maroun.abi-ghanem@univ-lyon1.fr}

\affiliation{Universite Claude Bernard Lyon 1, CNRS, Institut Lumière Matière, Villeurbanne, France}

\date{\today}

\begin{abstract}
We investigate the propagation of surface acoustic waves (SAWs) in a layered half-space system comprising a continuous, sub-wavelength-thick layer weakly adhering to a substrate. Using finite element simulations, we demonstrate that this configuration—without requiring surface structuration—gives rise to frequency ranges bounded in $k$-space and characterized by strong SAW attenuation, which we term adhesion-induced resonant attenuation zones. We show that these attenuation zones closely mimic the resonant behavior typically observed in locally resonant metamaterials and can be understood through a mass-spring analogy, where the adhesion between the layer and substrate governs the frequency and width of the attenuation zones. As a practical demonstration, we propose a  bilayer  configuration  as  a  practical route  to  experimentally  realize  adhesion-induced  resonant attenuation of SAWs, where a soft and thin interfacial film serves as an intermediate adhesive bonding between the layer and substrate, providing a realistic and tunable interfacial stiffness. Our findings offer a simplified route to achieving SAW manipulation through continuous layered media with tunable adhesion, providing a practical alternative to complex structural designs in SAW-based devices across a broad frequency range.
\end{abstract}

\maketitle

\section*{Introduction}

\begin{figure}
\includegraphics{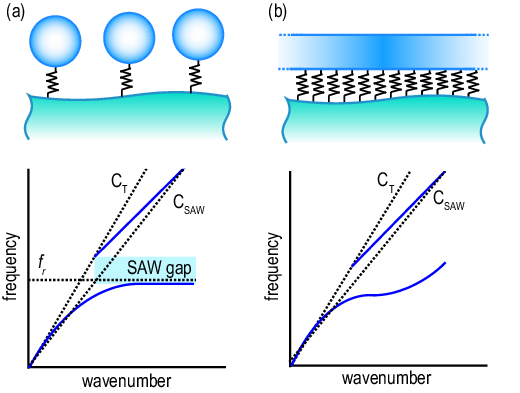}
\caption{\label{fig:fig1} Dispersive behavior of a SAW interacting with (a) isolated resonant elements characterized by a resonance frequency $f_r$, and (b) a thin elastic layer weakly adhering to the substrate via an interfacial stiffness $K_n$.}
\end{figure}

Surface acoustic waves (SAWs) are mechanical waves confined near the surface of elastic solids, with their energy predominantly localized within approximately one wavelength of depth \cite{royer1999elastic}. Their propagation can be significantly altered through attenuation zones, also referred to as avoided crossing bandgaps or SAW gaps, and arising from the hybridization of a propagating SAW mode with a locally resonant mode present within or coupled to the substrate. This hybridization causes a repulsive interaction between the dispersion of both modes, forming hybridized branches that exhibit characteristics of both the surface wave and the local resonance. These bandgaps are typically achieved using subwavelength-scale elements \cite{yudistira2016nanoscale} or engineered defects \cite{ash2021subwavelength} that act as discrete resonators. Such resonators introduce sharp spectral features and enable tunable, frequency-selective control over SAW transmission and dispersion. The resulting SAW gap is generally bounded by two branches: an upper branch that begins at the crossing with the transverse wave branch (i.e. at a frequency $\omega = c_T.k$, where $c_T$ is the transverse wave speed in the substrate) and approaches the SAW wave speed at higher wavenumbers, and a lower branch capped by a horizontal asymptote corresponding to the resonator's resonance frequency (see Figure \ref{fig:fig1}(a)). This gap does not span the entire $k$-space but rather begins at the $c_T$ branch. Depending on the coupling strength and system configuration, these gaps can manifest as weak attenuation zones in the SAW dispersion \cite{geslain2016spatial}, or as broad regions of reduced SAW transmission \cite{eliason2016resonant}. 

Avoided crossing-induced SAW gaps provide fine control over wave dispersion \cite{oudich2018rayleigh}, attenuation \cite{ash2017highly}, and confinement phenomena \cite{benchabane2017surface}, and are therefore central to a wide range of surface acoustic phonon systems, ranging from phononic crystals \cite{jin2021physics,vasileiadis2021progress} to hybrid phonon-magnon \cite{matsumoto2024magnon} and phonon-polariton \cite{yudistira2014monolithic} coupling platforms. Eigenmode vibrations of discrete micro/nanofabricated pillars, typically an order of magnitude smaller in width than the acoustic wavelength, have traditionally been employed to induce such attenuation zones in the MHz and GHz regimes \cite{sledzinska20202d}. Beyond isolated pillars, these attenuation zones have also been observed in systems where SAWs interact with contact resonances of microsphere monolayers \cite{boechler2013,hiraiwa2015}. Unlike isolated resonant pillars, microspheres within these monolayers are mechanically coupled to their adjacent neighbors through adhesive contact interactions \cite{hiraiwa2015,otsuka2018}. When sphere-to-sphere coupling becomes significant, the lower SAW branch exhibits a plateau in the dispersion, followed by a re-emergent propagative behavior (\textit{i.e.}, a dispersion with positive slope) at higher wavenumbers \cite{vega2018contact,vega2017}. Interestingly, similar dispersive characteristics--a flat transition region followed by a renewed propagative branch--have also been observed in layered half-space media where a weakly adhering layer interacts with the underlying elastic half-space \cite{adh1,adh2,adh4}. Although such media do not display the classical resonator-induced SAW gap in $k$-space due to the re-emergent nature of the lower branch (see Figure \ref{fig:fig1}(b) for illustration), the amplitude associated with the lower branch becomes noticeably weaker than that of the upper branch for a given wavenumber past the plateau region \cite{adh2,adh4}. This reduced amplitude indicates that continuous adhesive layers can give rise to attenuation zones marked by pronounced reductions in SAW transmission over specific $\omega$-$k$ regions. Despite this, the potential of such dispersive features as a phononic mechanism for controlling SAW propagation and attenuation remains largely unexplored. Harnessing these interactions could offer an alternative to the use of discrete vibrating elements and could enable effective SAW manipulation without the need for surface structuration.

In this work, we investigate SAW transmission through a layered half-space medium comprising a continuous, sub-wavelength-thick layer weakly bonded to an elastic substrate. Using Finite Element (FE) analysis, we demonstrate that the SAW spectrum exhibits near-zero transmission within certain frequency intervals, which we identify as adhesion-induced resonant attenuation zones. These regions are bounded in $k$-space by distinct minimum and maximum $k$-values, and can be tuned by adjusting the interfacial stiffness--that is, the adhesion-- of the layer to the substrate. By comparing these transmission spectra to those of classical discrete sub-wavelength resonators, we establish a mass-spring analogy that allows us to define an adhesion-induced resonant frequency determined by the interfacial stiffness and the surface mass density of the layer. Our results show that resonant attenuation of SAWs can be achieved without resorting to patterned or structured surfaces. To further elucidate this, we compare our transmission spectra with dispersion curves, revealing that simply adjusting the adhesion of continuous layers is sufficient to induce attenuation zones that can drastically reduce SAW transmission. We also propose a practical design strategy for achieving such thin-film-induced resonant attenuation. Specifically, we introduce a bilayer configuration in which one layer behaves as a lumped mass and the other provides the effective interfacial stiffness, thereby demonstrating that our theoretical framework is directly applicable to realistic material systems. Our study offers a new approach for controlling SAWs through material bonding strategies, without the need for complex surface geometries. 

\section{Numerical modeling} \label{section1}

We employ FE analysis using COMSOL Multiphysics to carry out our numerical simulations. We model MHz SAWs with acoustic wavelengths ranging from 20 to 200 $\mu m$, propagating on a layered half-space consisting of a sub-wavelength thickness layer adhered to a semi-infinite substrate. Given that the minimum acoustic wavelength in this SAW range does not exceed 20 $\mu m$, we set the adhesive layer thickness $h = 1$ $\mu m$, to ensure the sub-wavelength regime in the thickness direction ($\sim \lambda/10$), and the substrate thickness to 10 $\lambda$ to approximate semi-infinite conditions. The materials for both the substrate and the layer are considered isotropic, modeled using their density ($\rho_m$), longitudinal wave velocity ($c_{L,m}$), and shear wave velocity ($c_{T,m}$), where the index "$m$" denotes the layer "$l$" or the substrate "$s$". We use mechanical properties found in microscale “slow-on-fast” layered half-space media \cite{slowonfast}, commonly used in SAW device technologies \cite{sawdevice1}. For this, we take $c_{T,l}/c_{T,s} < 1/\sqrt{2}$ and use glass-like properties for the substrate ($c_{L,s} = 5700$ m/s, $c_{T,s} = 3400$ m/s, $\rho_s = 2300$ kg/m$^3$) and polymer-like properties for the layer ($c_{L,l} = 3500$ m/s, $c_{T,l} = 2000$ m/s, $\rho_l = 1500$ kg/m$^3$). To model the adhesion between the layer and the substrate, we use the “thin elastic layer” boundary condition, approximating the interface as interfacial massless springs using a longitudinal surface stiffness named $K_n$ \cite{ktkn1,ktkn2}, inducing a discontinuity in the component of the particle displacement normal to the interface, giving the following boundary condition:

\begin{align} 
       \sigma_{22} = K_n (u_{2,s} - u_{2,l}).
   \label{eq:eq1}
\end{align}

In Equation \ref{eq:eq1}, the index $2$ denotes the vertical direction, following the $x_2$ axis (see Figure \ref{fig:fig2}(a)), $\sigma_{22}$ in the normal stress at the substrate-layer interface, and $u_{2,s}$, $u_{2,l}$ the normal particle displacements in the substrate and in the layer on both sides of the interface, respectively. We do not consider any effect of the adhesion on the tangential stress $\sigma_{12}$ and particle displacement $u_1$, where the index $1$ denote the horizontal direction, following the $x_1$ axis. This leads to a continuity of the stresses and displacements on both sides of the interface.
    
\subsection{Eigenfrequency study}
\label{subsection1}

\begin{figure*}
\includegraphics{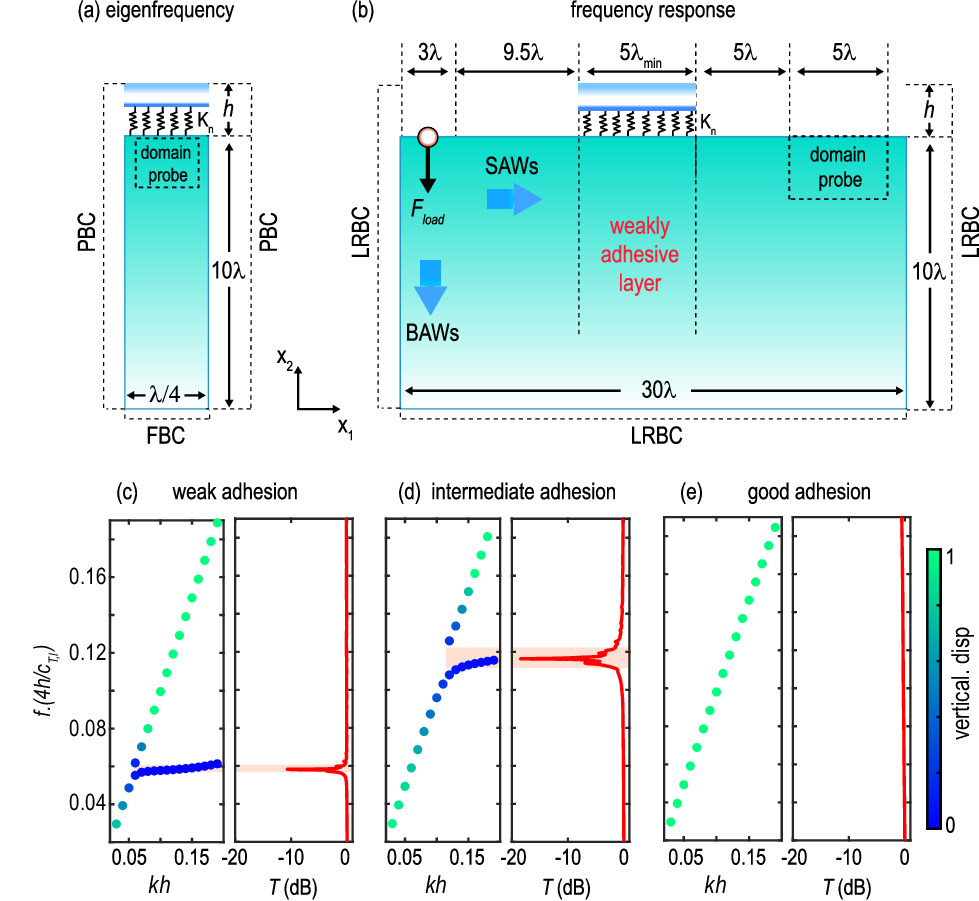}
\caption{\label{fig:fig2} (a-b) FE configurations used for the eigenfrequency and frequency response studies respectively. Dispersion curves and transmission spectra computed for (c) weak ($5 \times 10^{13}$ N/m$^3$), (d) intermediate ($2 \times 10^{14}$ N/m$^3$) and (e) good adhesion ($10^{16}$ N/m$^3$) of the layer to the substrate. The colorscale in the dispersion curves corresponds to the normalized maximum of the vertical displacement. In geometries (a-b), dimensions are expressed as a function of the SAW wavelength $\lambda$. PBC: Periodic Boundary Conditions, LRBC: Low-Reflecting Boundary Conditions. BAWs: Bulk Acoustic Waves}
\end{figure*}

We examine the dispersion curves of sagittal SAWs using an eigenfrequency analysis. To achieve this, we calculate the vibrational eigenmodes of the structure illustrated in Figure \ref{fig:fig2}(a), which represents a unit cell. The eigenfrequencies are determined by solving the elastodynamic equation of motion :

\begin{equation}
    \frac{\partial \sigma_{i,j,m}}{\partial x_j} + \rho_{m} \omega^2 u_{i,m} = 0;
\end{equation}
where the subscripts "$i$" and "$j$" correspond to the direction $x_1$ or $x_2$ respectively, and $\omega $ being the angular frequency.

We choose the width of the unit cell to be $\lambda/4$, allowing us to avoid fictitious folding on the dispersion curves due to the finite size of our unit cell. The meshing is configured to ensure a minimum of 15 nodes per wavelength, with a maximum of 20. 
We apply the Bloch-Floquet periodic boundary conditions (BCs) on the edges normal to the $x_1$-axis for both the substrate and the layer. We apply a low-reflecting boundary condition at the bottom of the substrate to simulate semi-infinite conditions. To obtain the dispersion curves, we treat the wavenumber as a free parameter, varying it between $k_{min} = 0.01$ rad/$\mu$m and $k_{max} = 0.5$ rad/$\mu$m. A domain probe is positioned at the substrate surface, with a width of $\lambda/8$ and a depth of $\lambda$ to measure the out-of-plane displacements. We plot the calculated dispersion curves based on the magnitude of the out-of-plane displacement of each mode.

\subsection{Frequency response study}

In conjunction with computing the SAW dispersion curves using an eigenfrequency analysis, we perform a frequency response study to compute the transmission spectra of SAWs propagating along the $x_1$ direction in a half-space with a locally weakly adhering layer. Figure \ref{fig:fig2}(b) shows the geometry of the domain used in our FE model. Following the approach outlined in \cite{femphononic}, we adapt the geometry and meshing based on the SAW wavelength in the substrate, expressed as $\lambda = c_{SAW}/f$, where $f$ is the frequency and $c_{SAW}$ is the surface wave velocity. The total width of the model is set to $30 \lambda$. We used a similar meshing configuration as the one used in the eigenfrequency study. 

We excite SAWs by applying a point load, located $1.5\lambda$ from the left edge of the domain, using a vertical force $F_{load}$ normal to the surface. To avoid reflections of bulk and surface waves in the domain, we apply low-reflecting boundary conditions to the left, bottom, and right edges of the model. The adhering layer is inserted at the center of the model over a width of $5 \lambda_{min}$, at the maximum frequency considered ($\lambda_{min} = c_{SAW}/f_{max}$), to ensure the layer is not wider than the substrate and to keep the same width for all the frequencies. The adhesion is simulated using the ``thin elastic layer" boundary condition, as described in Equation \ref{eq:eq1}. We perform a parametric frequency response analysis over a specific frequency range, from 10 MHz to 100 
MHz, following the harmonic equation of motion \cite{femphononic}:

\begin{equation}
    \frac{\partial \sigma_{i,j,m}}{\partial x_j} + \rho_{m} \omega^2 u_{i,m} = F_{load} \delta_{i2} e^{j \phi},
\end{equation}
$F_{load}$ being the harmonic excitation force at the angular frequency $\omega$ with a phase $\phi$ and $\delta$ is the Kronecker delta, since the force is only applied in the direction $x_2$.

To process the transmission spectrum (Figure \ref{fig:fig2}(c-e)), the magnitude of the vertical particle displacement is obtained using a $5\lambda$-wide and $1\lambda$-deep domain probe, located at a distance $2.5 \lambda$ from the right edge of the structure. We extract the transmission power by calculating the squared ratio of the out-of-plane displacement's magnitude with the weakly adhesive layer to the one without the layer.

\subsection{Results}

Figures \ref{fig:fig2}(c-e) depict the dispersion curves and transmission spectra of SAW propagation based on the FE models illustrated in Figures \ref{fig:fig2}(a) and \ref{fig:fig2}(b), respectively, where the frequency axis is normalized with respect to the fundamental mode of transverse vibration of the film $c_{T,l}/4h$. The adhesion strength between the layer and the substrate was systematically varied by adjusting the normal interfacial stiffness $K_n$. The stiffness values ranged from $5 \times 10^{13}$ N/m$^3$ (the weak adhesion case) to $10^{16}$ N/m$^3$ (the good adhesion case), with an intermediate value of $2 \times 10^{14}$ N/m$^3$ \cite{adh1}. For the good adhesion case (Figure \ref{fig:fig2}(e)), we observe a non-dispersive SAW mode with a wave speed of around $3000$ m/s, which corresponds to the Rayleigh wave speed in the substrate. The linear behavior up to $f/(c_{T,l}/4h) = 0.2$ indicates that the well adhering layer is acoustically transparent within this frequency range, and the dispersive behavior stemming from Rayleigh-to-Sezawa mode transition \cite{rayleigh2,rayleigh3} only occurs at much higher frequencies. This also verifies the sub-wavelength nature of the adhesive layer utilized in this study. Additionally, the vertical displacement of this mode, calculated over a depth equal to one acoustic wavelength and represented in a color scale (normalized, linear scale), is uniform throughout the entire range. This uniformity indicates the absence of any adhesion-induced attenuation within this frequency range. The transmission spectrum further supports this, as it shows uniform transmission with a value of 1 across the frequency range, demonstrating a consistent transmission through the strongly adhering layer.

In contrast, the cases of weak and intermediate adhesion reveal a significantly different behavior, characterized by a more nuanced acoustic response. For these cases, the acoustic dispersion curves show that the Rayleigh wave branch splits into two distinct branches, accompanied by the emergence of an attenuation zone within a specific frequency range (indicated by a shaded red region on the plots). This behavior is clearly visible in the transmission spectra as well, where we can observe attenuation dips reaching up to -20 dB in magnitude. In the weak adhesion case, we see that the attenuation zone is bounded by an upper branch that propagates at the Rayleigh wave speed and a lower branch that plateaus over a finite range in $k$-space. After this plateau, the lower branch eventually becomes propagative again at higher wavenumbers. Furthermore, comparing the weak and intermediate adhesion cases, we observe that the central frequency and width of the attenuation zone increase as adhesion is increased, indicating that adhesion strength directly impacts the characteristics of this zone. Interestingly, the propagative portion of the lower branch past the plateau shows a negligible displacement component in the substrate, as is illustrated by the blue color in the dispersion curves, signifying that it exists within the attenuation zone without altering it. This observation suggests that adhesion-induced attenuation zones may function analogously to those caused by isolated locally resonant structures \cite{boechler2013,yudistira2016nanoscale}, despite the lower branch not exhibiting the typical asymptotic behavior seen in such media (see Figure \ref{fig:fig1}).

\section{Analogy with locally resonant metamaterials} \label{section2}

In the limiting case where the shear modulus of the adhesive layer vanishes the analogy with an array of uncoupled mass-spring resonators becomes expected: the film can be regarded as a collection of non-interacting vertical oscillators, each behaving as a local resonant element. To further elucidate this, we compare the dispersion and transmission spectra of SAWs propagating on a surface with a weakly adhesive layer to those interacting with isolated, locally resonant oscillators. These oscillators are numerically modeled as point masses $m_r$ resting on a spring of linear stiffness $q_r$. Each mass is constrained to vibrate solely in its axial resonant mode, thus acting as single-degree-of-freedom oscillator characterized by a resonance frequency $f_r=1/(2\pi) \sqrt{q_r/m_r}$. We consider each resonator occupies a square surface area with side length $d_r$, which allows to express an effective interfacial stiffness of $K_r$ = $q_r$.$A_r$, where $A_r$ = $1/{d_r}^2$  for these spatially distributed oscillators.

\subsection{SAW interaction: local resonators vs adhesive layer}

\begin{figure}[h]
\includegraphics{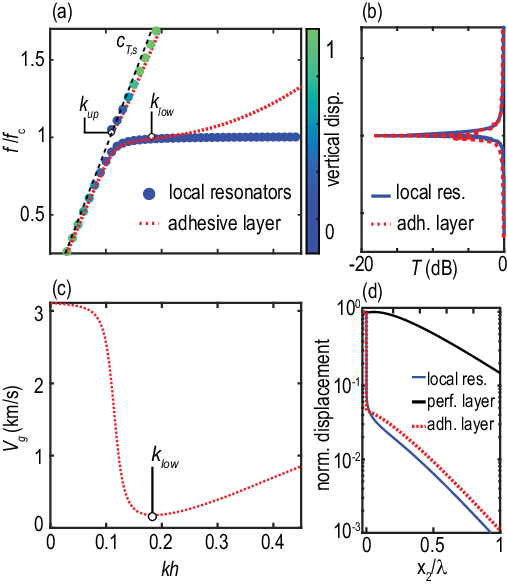}
\caption{\label{fig:fig3} Comparison between local resonators and adhesive layer through their (a) dispersion curves and (b) transmission spectra. The frequency scale is normalized by the frequency corresponding to the minimum of the transmission spectrum. The lower and upper boundary of the attenuation zone are represented in (a) by $k_{low}$ and $k_{up}$. (c) Variation of the group velocity of the lower branch as a function of $kh$, showing a minimum at $k_{low}$. (d) Depth profiles of the vertical displacement at $kh$ = 0.17 with $x_2 = 0$ at the substrate's surface, $x_2 > 0$ being in the substrate and $x_2 < 0$ in the layer/resonator. The displacement is normalized by the maximum of the displacement in the depth (located in the layer or in the resonator).}
\end{figure}

Figures \ref{fig:fig3}(a-b) show the SAW dispersion and transmission spectra in the presence of local resonators (blue markers and blue spectrum). The parameters ($q_r$ and $m_r$) used to model the local resonators are chosen such that the resonant frequency $f_r$ matches the frequency dip observed in the intermediate adhesion case in Figure \ref{fig:fig2}(d). In both figures, the frequency is normalized relative to the central frequency of the transmission dip $f_c$. In the dispersion curve (Figure \ref{fig:fig3}(a)), we see a classic avoided crossing behavior between the SAW and the local resonance of the spring-mass system, with the lower and upper branches behaving as described earlier in Fig. \ref{fig:fig1}(a). The displacement amplitude of both these branches near the resonant frequency approaches zero, giving rise to a SAW attenuation zone centered around this frequency. This zone is visible in the transmission spectrum as a dip reaching up to -20 dB (see Figure \ref{fig:fig3}(b)).  

We now contrast the dispersion obtained from the weakly adhesive layered-half-space model (shown in Figure \ref{fig:fig2}(d)), to that obtained from the locally resonant model. We used the central frequency of the attenuation zone of Figure \ref{fig:fig3}(b) to normalize the frequency. When comparing both the adhesive layer model (red) and the local resonators model (blue), we first observe that the upper branches of both models exhibit similar behaviors, both behaving as a Rayleigh wave. In addition, we see that the lower branches of both models behave qualitatively similarly at low to mid $k$, but notable differences start to emerge at high $k$. More specifically, we notice that the lower branch of the adhesive layer model exhibits an inflection point around $kh$ = 0.2. Below this inflection point, the behavior of both models align closely; however, beyond this point, the lower branch of the adhesive layer model tends asymptotically toward the fundamental flexural mode (\textit{i.e.} Lamb mode) of the layer \cite{adh1,adh2}, ultimately becoming propagative again. Despite this notable divergence in the dispersive behavior of the lower branches at high $k$, a comparison of the transmission spectra (see Figure \ref{fig:fig3}(b)) from both models reveals a striking resemblance in their shapes. This observation suggests that both systems behave analogously in terms of their SAW transmission characteristics and indicates that the behavior of the lower branch beyond the inflection point does not significantly influence this SAW transmission. As a result, the weakly adhesive layer can be viewed as a mass-spring local resonator over a certain range of wavenumbers, below the inflection point. 

To empirically establish the range of validity for this analogy, we define four key quantities for the adhesive layer case: $k_{low}$, $k_{up}$, $f_{low}$ and $f_{up}$, which represent the lower and upper boundaries for wavenumbers and frequencies, respectively (see Figure \ref{fig:fig3}(a)). The lower boundary, corresponding to where the first mode is nearly flat, is set at the minimum group
velocity, $V_g$, of the first mode, as shown in Figure \ref{fig:fig3}(c). The upper boundary is defined at the onset of the second mode, identified by its intersection with the shear wave velocity in the substrate, $c_{T,s}$ \cite{otsuka2018,slowonfast}. Within this $k$-range, defined as $k_{up}$ $<$ $k$ $<$ $k_{low}$, we identify an adhesion-induced resonant attenuation zone centered around $f_c$ $=$ $(f_{up}$+$f_{low})/2$ and having a width $\Delta f$ $=$ $f_{up}$-$f_{low}$. Unlike classical locally resonant SAW attenuation zones, which are unbounded at high $k$ due to the asymptotic behavior of the locally resonant flat branch, this adhesion-induced attenuation zone is bounded in $k$-space. This constraint occurs because the lower branch does not remain flat at high $k$-values but instead becomes propagative, approaching the fundamental flexural mode of the adhesive layer. Despite this difference, this zone produces sharp attenuation dips in the SAW transmission spectra, effectively mimicking the behavior of SAW gaps defined by flat branches extending across the entire $k$-range.

The analogous behavior of SAWs interacting with an adhesive layer and local resonators is further illustrated by examining the depth profile of the normal displacement (Figure \ref{fig:fig3}(d)) at $kh$ = 0.17, just below $k_{low}$, with $x_2 \ge 0$ corresponding to the displacement in the substrate and $x_2 < 0$ the displacement in the layer/resonator. In the case of a perfectly adhesive layer, the displacement profile is continuous throughout the structure, extending to a depth of approximately one wavelength. In contrast, when a weakly adhesive layer is present, minimal displacement reaches the substrate due to the discontinuity in the displacement profile over the depth. Structures incorporating local resonators exhibit comparable minimal acoustic displacement in the substrate, that is about three orders of magnitude smaller than the perfectly adhering layer. This is highlighted in red in Figure \ref{fig:fig3}(d), representing the displacement in the substrate. This shows that near $k_{low}$, the acoustic energy is mainly confined within the resonator - whether it be a discrete resonator or an adhesive layer - highlighting that the adhesive layer can function analogously to local resonators.

\subsection{Quantitative analysis and limits of the analogy}

\begin{figure}
\includegraphics[scale=1]{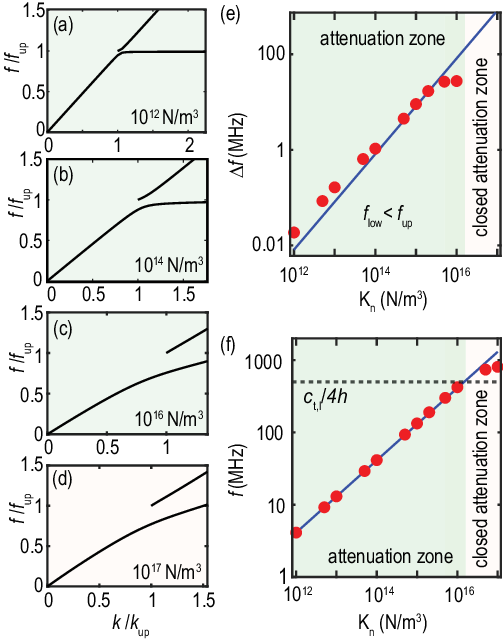}
\caption{\label{fig:fig4} (a-d) Dispersion curves for different values of $K_n$, normalized by the upper boundary of the attenuation zone, $f_{up}$ and $k_{up}$. The lower branch is plotted only up to $k_{low}$, to highlight the range of validity of the analogy with mass-spring resonators. The green background in (a-c) denotes the presence of an attenuation zone in the dispersion curve. The light red background in (d) indicates the closure of the attenuation zone. (e-f) Variation of $\Delta f$ and $f_{c}$ as a function of the interfacial stiffness, respectively.}
\end{figure}

To analyze the behavior of the adhesion-induced attenuation zone as a function of adhesion, we compute dispersion curves for a range of normal interfacial stiffnesses $K_n$, spanning five orders of magnitudes. Specifically, we consider values from $K_n = 10^{12}$ N/m$^3$ (very weak adhesion) to $K_n = 10^{17}$ N/m$^3$ (strong adhesion). Representative dispersion curves for four adhesion cases ($K_n = 10^{12},10^{14},10^{16}, 10^{17}$ N/m$^3$) are shown in Figures \ref{fig:fig4} (a-d). Each dispersion curve is terminated at $k = k_{low}$ to emphasize the range of validity defined by $k_{low}$. In this range, the dispersion of the adhesive layer closely follows that of a single locally resonant mass-spring oscillator. To better illustrate the evolution of the attenuation zone across this wide adhesion range, we normalize the frequency by $f_{up}$ and the wavenumber by $k_{up}$, making all the upper branches start at the same non-dimensionalized frequency and wavenumbers. These figures show that, as adhesion increases, the $k$-range of the attenuation zone generally decreases (\textit{i.e.}, $k_{low}$ decreases) while its width increases. We note that an analytical layered half-space-model with intermediate adhesion yields similar dispersive behavior (see Supplemental Material, section S.1 \cite{SI, transfermatrix1, transfermatrix2, rayleigh1885, khanolkar2015apl}).

For each case, we extract the width and the central frequency of the attenuation zone, respectively named $\Delta f$ and $f_c$, and plot them as a function of $K_n$, as is shown in Figures \ref{fig:fig4}(e-f). We first see that $\Delta f$ initially increases with adhesion, reaching a peak around $K_n = 5.10^{15}$ N/m$^3$, before falling down and attaining a minimum near $K_n = 2.10^{16}$ N/m$^3$. Beyond this point, $\Delta f$ becomes negative (\textit{i.e.}, when $f_{low} > f_{up}$), signaling the closure of the attenuation zone. This behavior of $\Delta f$ allows to delimit the attenuation region (highlighted in green) across the studied adhesion range. Within this region, we see that the increasing trend in $\Delta f$ is accompanied by a concurrent increase in $f_c$, which rather follows a linear progression (in logarithmic scale) until the attenuation regions closes. Notably, both $\Delta f$ and $f_c$ exhibit a considerable rise spanning several orders of magnitude over the adhesion range explored.

To quantify the variation of $f_c$, we define an adhesion-induced resonant frequency, $f_A$, such that:

\begin{equation}
    f_A = \frac{1}{2\pi} \sqrt{\frac{K_n}{\rho_l h}}.
    \label{eq:eq4}
\end{equation}

In Figure \ref{fig:fig4}(f), we plot this variation in blue and observe that $f_A$ accurately describes the behavior of the $f_c$ as a function of $K_n$ within the attenuation region. Drawing from the analogy with local resonators \cite{eliason2016resonant,resonators5}, the adhesive layer can thus be interpreted as a one-dimensional mass-spring resonator in this region, with a resonant frequency equal to $f_A$. Comparing $f_A$ to the lowest thickness resonance mode of a fixed-free layer \cite{resonance}, given by $c_{T,l}/(4h)$, we find that $f_A$ is always less than $c_{T,l}/(4h)$. This observation is expected as $c_{T,l}/(4h)$ represents the cutoff frequency of the guided modes in the layer. Below this threshold, the layer acts as a lumped mass, rather than supporting propagating wave modes. Note that this analogy also holds when localized resonators are placed on top of a weakly bonded layer, leading to coupling between the two systems that resembles the interaction of coupled harmonic oscillators. This point is examined in more detail in the Supplemental Material, section S.2 \cite{SI}.

To describe the variation of $\Delta f$, we define the frequency width of the attenuation zone as $\Delta f_A$ also obtained from a locally resonant mass-spring model \cite{otsuka2018}:

\begin{equation}
    \Delta f_A = \frac{K_n}{4 \pi \rho_s c_{T,s}} \sqrt{1-\frac{c_{T,s}^2}{c_{L,s}^2}}.
    \label{eq:eq5}
\end{equation}
We plot $\Delta f_A$ in blue in Figure \ref{fig:fig4}(g). We find excellent agreement with $\Delta f$ for $f_A < c_{T,l}/(4h)$. This confirms that the analogy between adhesive layer and local resonators also extends to the width of the adhesion-induced attenuation zone. 

It is worth noting here that, in classical mass-spring resonators, the attenuation zone width is primarily controlled by the surface density of the resonators $A_r$, while the resonant frequency is governed by the spring constant $q_r$. In contrast, for continuous adhesive layers, both of these quantities are merged into the interfacial stiffness $K_n$ as an effective lumped parameter, the interfacial stiffness $K_n$, which simultaneously governs both the width and central frequency of the attenuation zone. Writing $K_n = K_a A_a$, where $K_a$ represents the local adhesion stiffness (analogous to $q_r$) and $A_a$ the surface density of adhesion points (analogous to $A_r$), enables the continuous interface to be viewed as an effective collection of discrete contact points. These quantities can be linked to physical features such as surface roughness, the presence of asperities, or localized van der Waals forces \cite{rugo1,rugo2,rugo3,israelachvili2011intermolecular}, thereby reinforcing the analogy with classical mass-spring model of locally resonant systems.\\

\section{Bilayers for adhesion-induced resonant attenuation of SAWs}

\begin{figure}[h]
\includegraphics[scale=1]{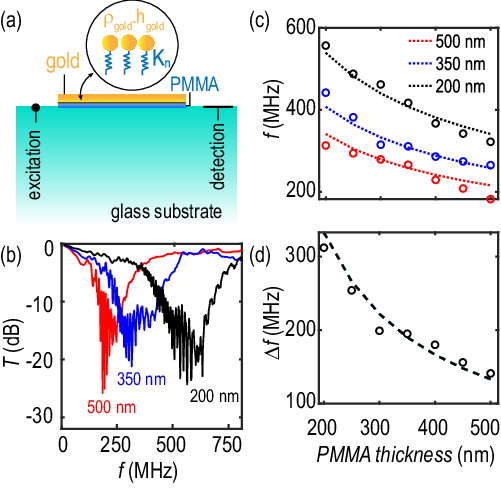}
\caption{\label{fig:fig5} FE frequency-response simulations on bilayer structures using a PMMA layer as an adhesive layer between a gold film and a glass substrate (a) Schematic of the simulated structure. Inset shows an analogous mass-spring resonator structure. Varying the PMMA thickness, $h_p$, which mimics changes in the interfacial stiffness $K_n$, modifies the central frequency and width of the attenuation zone are modified. The gold-layer thickness $h_g$ also influences the central frequency, as predicted in Equation \ref{eq:eq5}. (b) Transmission curves (in dB) for three cases, for which the thickness of the gold film and the PMMA layer are equal. In these cases, $h_g = h_p$ = 200nm, 350 nm and 500 nm. (c) Central frequency of the attenuation zone as a function of $h_{p}$ for different values of $h_g$, extracted from the position of the transmission minimum. (d) Width of the attenuation zone as a function of $h_p$, determined from the full width at half minimum. Open circles denote values obtained from the FE simulations, and dashed curves correspond to values given by the analytical expressions \ref{eq:eq4} and \ref{eq:eq5}.}
\end{figure}

We have now demonstrated that a weakly bonded layer can provide an effective and straightforward means to selectively attenuate the propagation of SAWs. To propose a practical implementation of a controlled adhesion between a layer and a substrate, we present a numerical, frequency-response study based on the model shown in Figure \ref{fig:fig5}(a), following the approach presented in Section \ref{section1}. More details about the model are available in Supplemental Material \cite{SI,pmmafilm2,pmmafilm3} in Section S.3. In our example, a thin gold layer is bonded to a glass substrate using a thin Poly-methyl methacrylate (PMMA) intermediate layer, which serves to mimic a controlled degradation of the adhesion between gold and glass. By varying the thickness of the interfacial PMMA layer, we can simulate different levels of effective adhesion between the gold film and the substrate, with the corresponding interfacial stiffness $K_n$ given by:
\begin{equation}
    K_n = c_{L,\mathrm{p}}^2\rho_{\mathrm{p}}/h_{\mathrm{p}},
    \label{eq:eq6}
\end{equation}
where, $h_p$, $c_{L,p}$ and $\rho_p$ denote the thickness, longitudinal wave velocity and density of the interfacial layer (PMMA), respectively.

Following Equation \ref{eq:eq6}, we first performed a parametric study in which the effective interfacial stiffness was reduced by varying the PMMA thickness $h_{p}$ from 200 to 500 nm \cite{pmmafilm}, which decreases $K_n$ from $4.4 \times 10^{16} ~N/m^3$ to $1.8 \times 10^{16} ~N/m^3$. Figure \ref{fig:fig5}(b) shows the transmission spectra of the gold/PMMA bilayer for three different PMMA and gold thicknesses ($h_{p} = h_g$ = 200, 350 and 500 nm). In all cases, we see the presence of an attenuation zone with a depth exceeding 20 dB. The central frequency of this zone shifts with the thickness of the PMMA, and the trend shows a very good agreement with the predicted resonance frequency $f_A$, calculated using Equations \ref{eq:eq4} and \ref{eq:eq6}. In Figure \ref{fig:fig5}(c), we plot the frequency of the transmission minimum as a function of the PMMA thickness for three different gold layer thicknesses $h_p$. We observe that the adhesion-induced attenuation zone follows the expected variation of $f_A$, demonstrating that it can be controlled in two independent ways: (i) increasing the PMMA thickness reduces $K_n$ (Equation \ref{eq:eq6}), corresponding to weaker adhesion and lower resonance frequency; and (ii) increasing the gold layer thickness increases the effective mass, which also lowers the resonance frequency (Equation \ref{eq:eq4}).

We also extracted the full width at half minimum of the transmission dips and plotted it as a function of the PMMA thickness in Figure \ref{fig:fig5}(d). The width decreases as the thickness of the PMMA layer increases, that is, as the effective adhesion of the gold layer to the substrate becomes weaker, in agreement with the trends observed in Figure \ref{fig:fig4}. The dashed curves in Figures \ref{fig:fig5}(c,d) correspond to the analytical predictions from Equations \ref{eq:eq4} and \ref{eq:eq5}, confirming that a simple bilayer configuration (such as the one proposed here) can thus be employed for adhesion-induced resonant attenuation of SAWs. Furthermore, we note that the results from the FE simulations agree well with the simplified analytical approach that assumes $K_t \rightarrow \infty$, even though the PMMA layer has a finite shear modulus, thus a finite value of $K_t$.

\section{Discussion and conclusion}

We have shown that a weakly adhesive layer can create attenuation zones in the transmission spectra of SAWs, akin to those observed in locally resonant metamaterials. By drawing an analogy between the adhesive layer and an array of local resonators, we modeled its behavior using a simple mass-spring resonator model. This analogy allowed us to identify an adhesion-induced resonant attenuation zone centered at a frequency $f_A$, which depends on the interfacial adhesion stiffness $K_n$, as well as the density and thickness of the layer. From a practical engineering standpoint, this approach provides a useful tool for adhesion assessment in the context of non-destructive evaluation \cite{adh4,adhcurve1,adhcurve2}, as it offers a simplified alternative to classical inverse problem methods used to derive dispersion curves to quantify interfacial stiffnesses.

We have proposed a bilayer configuration as a practical means to design an experimental scenario where adhesion drives local resonance. In our example, a gold layer is bonded to a glass substrate through a thin interfacial PMMA layer, allowing the effective adhesion to be tuned simply by varying the PMMA thickness. Our simulations show that the analytical model accurately predicts both the central frequency and the width of the attenuation zone, demonstrating that realistic constinuous structures can be designed to control SAWs using adhesive layers.

Although the adhesive layer in our study is continuous and lacks structuration, it exhibits a phononic feature, namely SAW resonant attenuation zone, that is classically associated with structured locally resonant surfaces. Previous studies have shown that, similarly, unstructured layered media can support unusual phononic features traditionally sought through complex surface or material structuration. This includes Dirac cones \cite{diracC1,diracC2}, backward-propagating waves \cite{resonance,backwards} and wave focusing \cite{focusing}. Our results align with this growing body of work, suggesting that by carefully tuning material and interfacial properties, rather than relying on structural designs, one can achieve comparable, if not superior, control over acoustic wave phenomena.

Our model is adaptable to a broad range of SAW-related applications and frequency regimes where interfacial mechanics play a central role \cite{adh3}. This includes low-frequency domains such as seismic shielding (Hz range) \cite{earth2,earth3}, ultrasonic regimes like MHz-range SAW filters where interfacial properties can be dynamically tuned through electrostatic \cite{adh4} or optical methods \cite{microlensing}, and hypersonic regimes (GHz and sub-THz), where weakly adhered van der Waals materials \cite{adh5} or complex oxides \cite{adh6} can serve as both passive and active components for SAW-related phonon engineering.

Finally, our analysis reveals that adhesion-induced resonant attenuation of SAWs does not depend on the intrinsic mechanical properties of the adhesive layer (specifically its longitudinal and transverse sound speeds) but is primarily governed by its interfacial mechanics (mainly the interfacial stiffness and surface mass density). This opens the door to the use of mechanically heterogeneous materials, such as those found in biological media \cite{sensing3, nacre}, in the design of locally-resonant metamaterials for SAWs. Overall, our work presents an alternative route for manipulating SAWs, offering a simpler route to achieving resonant attenuation zones without the need for complex structural architectures. 

\section{Acknowledgements}
T.D. acknowledges financial support from Agence Nationale de la Recherche (ANR) through grant funding ANR-20-CE43-0004. M.A.G. acknowledges financial support from ANR through grant funding ANR-22-CE05-0001.

\bibliography{apssamp}

\clearpage

\section{Supplemental Material}

\renewcommand{\thesection}{S.\arabic{section}}

\subsection{Analytical model}

\label{section1SI}

The model used to calculate the analytical dispersion curves, presented in Figure \ref{fig:figSI}(b-c) in Section \ref{section2SI}, is detailed here. These calculations are based on the Transfer Matrix formalism from Thomson and Haskell \cite{transfermatrix1,transfermatrix2}.

\subsubsection{Transfer Matrix formalism for SAWs propagating in layered media}

We consider Surface Acoustic Waves (SAWs) propagating in the $x_1$ direction at the surface of a structure composed by a semi-infinite substrate topped with an adhesive layer, as is presented in Figure 2(a) of the main text. $x_2$ is oriented in the direction normal to the substrate's surface and pointing downwards (opposite direction to Figure 2(a) of the main text). The equation of propagation for elastic waves in solid media is expressed as follows :

\begin{equation}
\rho_m \frac{\partial^2 u_{i,m}}{\partial t^2} = c_{ijko,m} \frac{\partial ^2 u_{o,m}}{\partial x_k \partial x_j},
\label{eqn:eq0a1}
\end{equation}

where the subscript $m$ denotes the propagation in the medium $m$ ($s$ for the substrate and $l$ for the layer), $u_{i,m}$ and $u_{o,m}$ the particle displacements along the direction $i$ and $o$, and $c_{ijko,m}$ the component of the elasticity constants' tensor for the medium $m$ . Considering only the saggital plane ($x_1$, $x_2$ plane), isotropic media and the symmetry of the tensor, $c_{ijko,m}$ can be simplified into 4 components with only 2 independent values: $c_{1111,m} = c_{2222,m} = \lambda_m + 2 \mu_m$, $c_{1122,m} = \lambda_m$ and $c_{1212,m} = \mu_m$, where $\lambda_m$ and $\mu_m$ are the Lamé constants of the medium $m$ \cite{royer1999elastic,rayleigh1885}.

Solving Equation \ref{eqn:eq0a1}, the expressions of the tangential and normal displacements, referred to respectively as $u_{1,m}$, $u_{2,m}$, are obtained:

\begin{widetext}
    \begin{equation}
        \begin{cases}
            u_{1,m} = [A_{1,m} e^{-jP_m}+A_{2,m}e^{+jP_m} + \beta_{2,m}(A_{3,m}e^{-jQ_m}-A_{4,m}e^{+jQ_m})] e^{j(\omega t-k x_1)}\\
            u_{2,m} = [\beta_{1,m}(A_{1,m} e^{-jP_m}-A_{2,m}e^{+jP_m}) - A_{3,m}e^{-jQ_m}-A_{4,m}e^{+jkQ_mx_2}] e^{j(\omega t-k x_1)}
        \end{cases},
        \label{eqn:eq0a2}
    \end{equation}
\end{widetext}

where $A_{1,m}$, $A_{3,m}$ are the amplitudes of the incident longitudinal and shear waves, propagating towards $x_2 +$, respectively, and $A_{2,m}$ and $A_{4,m}$ the backward longitudinal and shear waves, propagating towards $x_2 -$, respectively, $\omega$ is the pulsation of the surface wave and $k$ its wavenumber along the direction $x_1$. The quantity $\beta_{1,m}$, $\beta_{2,m}$, $P_m$ and $Q_m$ are associated to the wavenumber of the SAWs along $x_2$ for the longitudinal and tangential component of the wave. They are expressed as follows:

\begin{equation}
    \begin{cases}
        \beta_{1,m} = -j\sqrt{1-\frac{V_\phi^2}{c_{L,m}^2}} \\
        \beta_{2,m} = -j\sqrt{1-\frac{V_\phi^2}{c_{T,m}^2}} \\
        P_m = k\beta_{1,m}x_2 \\
        Q_m = k\beta_{2,m}x_2
    \end{cases}
    \label{eqn:eq0a3}
\end{equation}

$V_\phi$ being the phase velocity of the SAWs in the structure. From the solutions of the wave equation (Equation \ref{eqn:eq0a2}), the elastodynamic equation of motion in continuous solid media without external volumic forces leads to the expression of the normal and tangential stresses, respectively $\sigma_{22,m}$ and $\sigma_{12,m}$, in the medium $m$:

\begin{equation}
    \begin{cases}
        \frac{\partial \sigma_{ij,m}}{\partial x_j} = \rho_m \frac{\partial ^2 u_{i,m}}{\partial t^2}\\
    \end{cases}
    \label{eqn:eq0a4}.
\end{equation}

Using Equation \ref{eqn:eq0a2} and solving Equation \ref{eqn:eq0a4}, the displacements and stresses can be expressed as follows in the transfer matrix formalism \cite{transfermatrix1,transfermatrix2}:

\begin{equation}
\begin{bmatrix}
u_{1,m} \\
u_{2,m} \\
\sigma_{12,m} \\
\sigma_{22,m}
\end{bmatrix}
= 
[M_m(x_2)]
\begin{bmatrix}

    A_{1,m} + A_{2,m}\\
    A_{1,m} - A_{2,m}\\
    A_{3,m} + A_{4,m}\\
    A_{3,m} - A_{4,m}
    
\end{bmatrix}
e^{j(\omega t-kx_1)},
\label{eqn:eq1a}
\end{equation}

where $[M_m(x_2)]$ is the matrix as follow:

\begin{widetext}
    \begin{equation}
        \begin{bmatrix} 
            \cos(P_m) & -j\sin(P_m) & -j\beta_{2,m} \sin(Q_m) & \beta_{2,m} \cos(Q_m) \\
            -j\beta_{1,m} \sin(P_m) & \beta_{1,m} \cos(P_m) & -\cos(Q_m) & j\sin(Q_m) \\
            -2k\beta_{1,m} \sin(P_m) \rho_m c_{T,m}^2 & -2jk\beta_{1,m} \cos(P_m) \rho_m c_{T,m}^2 & -jk[\beta_{2,m}^2-1]\cos(Q_m)\rho_m c_{T,m}^2 & -k[\beta_{2,m}^2-1]\sin(Q_m)\rho_m c_{T,m}^2 \\
            -jk[\beta_{2,m}^2-1]\cos(P_m)\rho_m c_{T,m}^2 & -k[\beta_{2,m}^2-1]\sin(P_m)\rho_m c_{T,m}^2 & 2k\beta_{2,m} \sin(Q_m) \rho_m c_{T,m}^2 & 2 j k \beta_{2,m} \cos(Q_m) \rho_m c_{T,m}^2
        \end{bmatrix}.
        \label{eqn:eq2a}
    \end{equation}
\end{widetext}

\subsubsection{Boundary conditions for adhesive layer}

Since the substrate is assumed semi-infinite along the $+ x_2$ direction, no wave propagates backwards according to the Sommerfeld conditions and $A_{2,s} = A_{4,s} = 0$.

The adhesion of the layer to the substrate is modeled using normal interfacial stiffness, respectively $K_n$ \cite{ktkn1,ktkn2}. 

Considering the surface of the layer at $x_2 = 0$ and the interface between the layer and the substrate at $x_2 = h$, the boundary conditions are the following:

\begin{equation}
    \begin{cases}
        \sigma_{12,l} (x_2 = h) = \sigma_{12,s} (x_2 = h) \\
        \sigma_{22,l} (x_2 = h) = \sigma_{22,s} (x_2 = h) \\
        u_{1,s}(x_2 = h) =  u_{1,l}(x_2 = h) \\
        \sigma_{22,l}(x_2 = h) =  K_n \left( u_{2,s}(x_2 = h) - u_{2,l}(x_2 = h) \right)\\
        \sigma_{12,l}(x_2 = 0) = 0 \\
        \sigma_{22,l}(x_2 = 0) = 0
    \end{cases}
    .
    \label{eqn:eq4a}
\end{equation}

The boundary conditions in $x_2 = h$ gives rise to the following expression :

\begin{equation}
    \begin{bmatrix}

        u_{1,l}(x_2 = h)\\
        u_{2,l}(x_2 = h)\\
        \sigma_{12,l}(x_2 = h)\\
        \sigma_{22,l}(x_2 = h)
    
    \end{bmatrix}
=[M_{int}]
    \begin{bmatrix}

        u_{1,s}(x_2 = h)\\
        u_{2,s}(x_2 = h)\\
        \sigma_{12,s}(x_2 = h)\\
        \sigma_{22,s}(x_2 = h)
    
    \end{bmatrix}
    ,
    \label{eqn:eq5a}
\end{equation}

with $[M_{int}]$ the interfacial stiffnesses matrix such that:

\begin{equation}
[M_{int}] =
        \begin{bmatrix}
        1 & 0 & 0 & 0 \\
        0 & 1 & 0 & -\frac{1}{K_n} \\
        0 & 0 & 1 & 0 \\
        0 & 0 & 0 & 1
    \end{bmatrix}
    .
    \label{eqn:eq6a}
\end{equation}

The boundary conditions in $x_2 = 0$ leads to:

\begin{equation}
    \begin{bmatrix}

        u_{1,l}(x_2 = 0)\\
        u_{2,l}(x_2 = 0)\\
        0\\
        0
    
    \end{bmatrix}
= [M_l(0)]
\begin{bmatrix}
    
        A_{1,l} + A_{2,l}\\
        A_{1,l} - A_{2,l}\\
        A_{3,l} + A_{4,l}\\
        A_{3,l} - A_{4,l}
        
    \end{bmatrix}
    e^{j(\omega t - k x_1)}
    . \label{eqn:eq7a}
\end{equation}

\subsubsection{Dispersion relation for SAWs}

By replacing the amplitudes in the layer in Equation \ref{eqn:eq7a}, by the amplitudes in the substrate using Equation \ref{eqn:eq5a}, the following expression is obtained :

\begin{equation}
    \begin{bmatrix}
    
        A_{1s}\\
        A_{1s}\\
        A_{3s}\\
        A_{3s}
        
    \end{bmatrix}
    e^{j(\omega t - k x_1)}
    =
    [M_{tot}]
    \begin{bmatrix}

        u_{1l}(x_2 = 0)\\
        u_{2l}(x_2 = 0)\\
        0\\
        0
    
    \end{bmatrix}
    , 
    \label{eqn:eq8a}
\end{equation}

$[M_{tot}]$ being :

\begin{equation}
    [M_{tot}] = [M_s(h)]^{-1}[M_{int}]^{-1}[M_l(h)][M_l(0)]^{-1}.
    \label{eqn:eq9a}
\end{equation}

Equation \ref{eqn:eq8a} leads to the dispersion equation by eliminating the terms $A_{1,s}, A_{3,s}, u_{1,l}(x_2 = 0)$ and $u_{2,l}(x_2 = 0)$. The components of the line $i$ and column $j$ of the matrix $[M_{tot}]$ being $M_{tot}^{i,j}$, the dispersion equation is :

\begin{equation}
    \frac{M_{tot}^{2,2} - M_{tot}^{1,2}}{M_{tot}^{1,1}-M_{tot}^{2,1}} - \frac{M_{tot}^{4,2}-M_{tot}^{3,2}}{M_{tot}^{3,1}-M_{tot}^{4,1}} = 0.
    \label{eqn:eq10a}
\end{equation}

Solving this equation leads to the dispersion curves of the layered half-space medium with intermediate adhesion.

\subsubsection{SAWs interacting with resonators}

If individual mass-spring resonators are now placed on top of the structure, such as in Figure \ref{fig:figSI}(a), the presence of the resonators modifies the boundary conditions at the surface of the layer at $x_2 = 0$ (Equation \ref{eqn:eq4a}). The normal stress is no more equal to zero  \cite{boechler2013,khanolkar2015apl}:

\begin{equation}
    \sigma_{22,l} (x_2 = 0) =  q_r A_r(u_{2,l}(x_2 = 0) - Z_r),
     \label{eqn:eq11a}
\end{equation}

with $Z_r$ the motion of the resonators' mass under the SAWs vibration. Assuming a harmonic motion with an amplitude $Z_{0r}$, $Z_r = Z_{0r} e^{j(\omega t - k x_1)}$, the equation of motion of the mass can be written :

\begin{equation}
    -\omega^2 m_r Z_r + q_r (Z_r-u_{2,l}(x_2 = 0)) = 0.
     \label{eqn:eq12a}
\end{equation}

This leads to the following expressions of $u_{2,l} (x_2 = 0)$ and $\sigma_{22,l} (x_2 = 0)$ :

\begin{equation}
\begin{cases}
         u_{2,l} (x_2 = 0) = \left( 1 - \frac{\omega^2}{\omega_r^2} \right) Z_r  \\
     \sigma_{22,l} (x_2 = 0) = - \omega^2 m_r A_r Z_r
\end{cases}
,
 \label{eqn:eq13a}
\end{equation}

$\omega_r$ being the pulsation of the oscillator at its resonance frequency $f_r$ such as $\omega_r = \sqrt{\frac{q_r}{m_r}} = 2 \pi f_r$.\\

Following the procedure used to obtain Equation \ref{eqn:eq10a}, the following dispersion equation is found:

\begin{widetext}
    \begin{equation}
            \frac{\left( M_{tot}^{2,2} - M_{tot}^{1,2} \right) \left( 1 - \frac{\omega^2}{\omega_r^2} \right) + \left( M_{tot}^{2,4} - M_{tot}^{1,4} \right) \left(- \omega^2 m_r A_r Z_r\right)}{M_{tot}^{1,1}-M_{tot}^{2,1}} - \frac{\left(M_{tot}^{4,2}-M_{tot}^{3,2}\right) \left( 1 - \frac{\omega^2}{\omega_r^2} \right)+ \left( M_{tot}^{4,4}-M_{tot}^{3,4}\right)\left(-\omega^2 m_r A_r Z_r\right)}{M_{tot}^{3,1}-M_{tot}^{4,1}} = 0.
        \label{eqn:eq14a}
    \end{equation}
\end{widetext}

The solution of this dispersion equation gives the dispersion curves of a SAWs propagating on top of a structure covered by resonators, as presented in Figure \ref{fig:figSI}(b),(c).

\subsection{Coupling between a weakly adhering layer and locally resonant elements} 

\renewcommand{\thefigure}{S.\arabic{figure}}

\label{section2SI}

We examine the coupling between the adhesion-induced resonant frequency of an adhesive layer $f_A$ and the resonant frequency $f_r$ of local resonators deposited on top. To this end, we investigate SAW interaction in a system composed of a substrate with an adhesive layer, on which 1D mass-spring resonators are placed (Figure \ref{fig:figSI}(a)).

We employ both analytical and numerical models to derive and analyze the dispersion curves of this coupled system. The analytical model is described in Section \ref{section1SI} while the Finite Element (FE) numerical model follows the methodology described in section I.A of the main text. We recall that the uncoupled resonant frequency of the local resonators is given by $f_r \propto \sqrt{q_r/m_r}$, with $q_r$ and $m_r$ the linear stiffness and mass of the resonators. The adhesion of the layer is selected within the range where the analogy between adhesive layer and local resonance remains valid, i.e, for $f_A < f_0$.

We investigate the coupling behavior across different regimes, ranging from weak coupling, where $f_r$ and $f_A$ are far from each other, to strong coupling, where $f_r \approx f_A$. The corresponding dispersion curves are shown in Figures \ref{fig:figSI}(b),(c). In the weak coupling scenario, in our case we took $f_A = 5 f_r$ (see Figure \ref{fig:figSI}(b)), the interaction between both resonating systems is minimal, and the dispersion curve resembles a superposition of the independent dispersion relations of the adhesive layer and the resonators. In this regime, the two components can be regarded as decoupled, leading to a primary attenuation zone induced by the local resonators at or near $f_r$, and a adhesive resonant attenuation zone associated with the adhesive layer at $f_A$.

As $f_r$ approaches $f_A$ (see Figure \ref{fig:figSI}(c)), strong coupling effects become apparent. Rather than distinct attenuation zones at $f_r$ and $f_A$, two significantly shifted zones emerge at $f_{r_1}$ and $f_{r_2}$. This behavior indicates that the system's influence on the SAW dispersion can no longer be interpreted by treating the adhesive layer and resonators independently. In this case, the original attenuation zones around $f_r$ and $f_A$ are repelled and displaced to $f_{r_1} < f_r$ and $f_{r_2} > f_A$, respectively. The repelled attenuation zones' frequencies in the strong coupling regime can be predicted using a two-degree-of-freedom mass-spring system, wherein the adhesion frequency corresponds to the resonance frequency of one of the system prior to coupling. As such, $f_{r1}$ and $f_{r2}$ can be expressed as:

\begin{figure*}[h]
\includegraphics[width=1\linewidth]{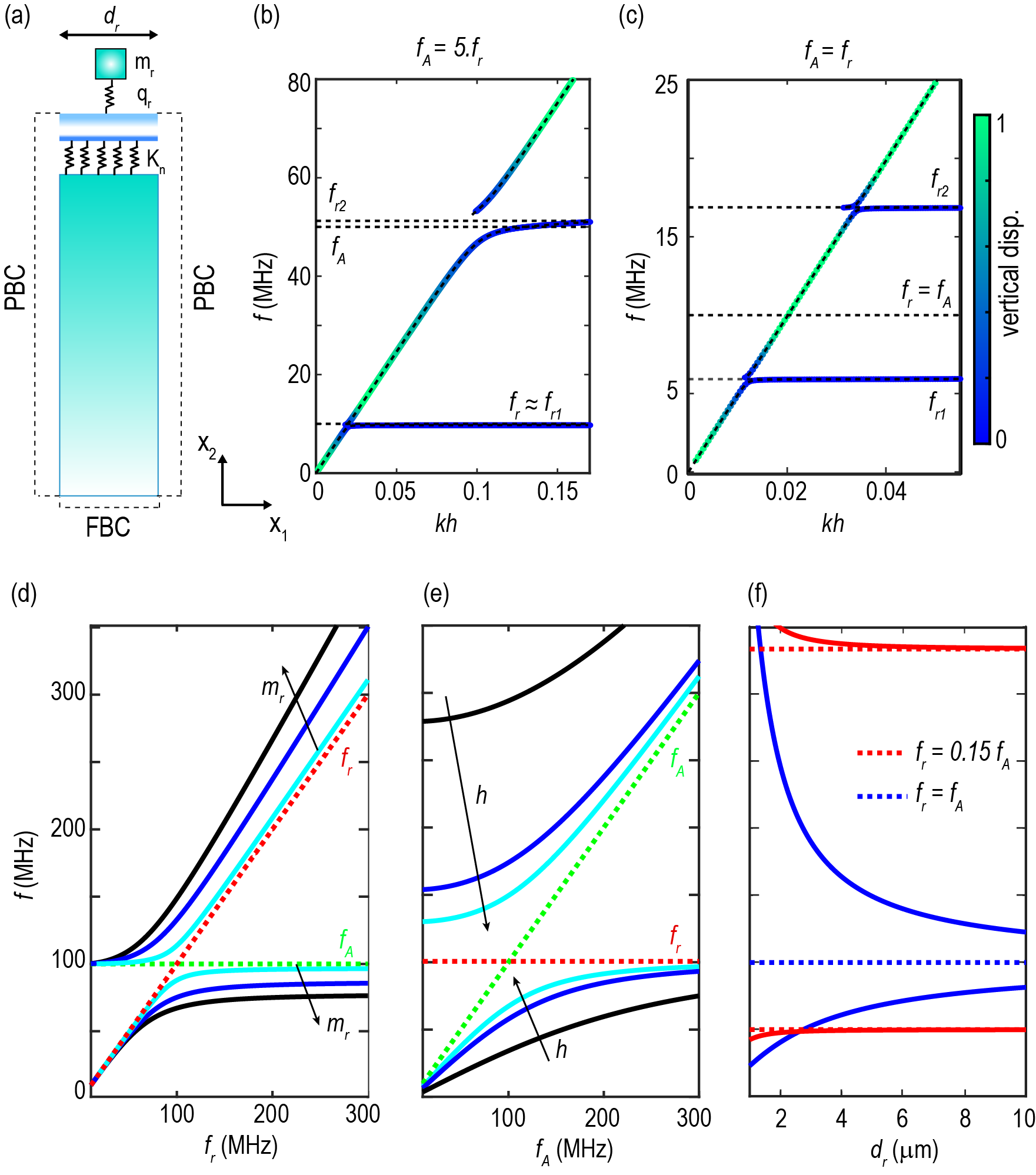}
\caption{\label{fig:figSI} (a) FE configuration used to compute the dispersion curves of a layered half-space model, including a weakly adhesive layer with spring-mass resonators placed on top of it. Computed dispersion curves for (b) $f_A = 5.f_r$ and (c) $f_A = f_r$. The dispersion curves are plotted up to $k_{low}$. Variation of the coupled frequencies $f_{r1}$ and $f_{r2}$ as a function of (d) $f_r$, (e) $f_A$, and $d_r$. The different colors in (d) show the influence of $m_r$. The different colors in (e) show the influence of $h$.} 
\end{figure*}

\begin{widetext}
    \begin{equation}
            \begin{cases}
                f_{r1} = \frac{1}{2\pi\sqrt{2}} \sqrt{\omega_A^2+\frac{q_r A_r}{\rho_l h}+ \omega_r^2 - \sqrt{(\omega_A^2+\frac{q_r A_r}{\rho h}+\omega_r^2)^2-4\omega_A^2\omega_r^2}}\\
                f_{r2} = \frac{1}{2\pi\sqrt{2}} \sqrt{\omega_A^2+\frac{q_r A_r}{\rho_l h}+ \omega_r^2 + \sqrt{(\omega_A^2+\frac{q_r A_r}{\rho h}+\omega_r^2)^2-4\omega_A^2\omega_r^2}}
            \end{cases}
            .
            \label{eq:eq15a}
    \end{equation}
\end{widetext}
 Equation \ref{eq:eq15a} appropriately captures the interaction between the adhesive layer and the local resonators within the frequency range where $f_A < c_{T,l}/4h$. To better understand the strength of this coupling, we plot in Figure \ref{fig:figSI}(d) the variation of the coupled resonances $f_{r_1}$ and $f_{r_2}$ as function of the uncoupled resonator $f_{r}$, while keeping a fixed adhesion frequency $f_{A}$. In the regime where $f_{r}$ $<<$ $f_{A}$, the lower branch $f_{r_1}$ closely follows $f_{r}$ (shown in dashed-red curve), while the upper branch $f_{r_2}$ remains close to $f_{A}$ (shown in dashed-green curve), showing that both resonances are decoupled. However, as $f_{r}$ approaches $f_{A}$, the two branches interact more strongly, leading to an avoided crossing. In this transition region, both $f_{r_1}$ and $f_{r_2}$ deviate from their original uncoupled values, although $f_{r_1}$ remains below and bounded by $f_{A}$.

This coupling behavior also depends on the resonator mass $m_{r}$. For smaller mass values, the branches tend to stay closer to the uncoupled frequencies $f_{r}$ and $f_{A}$ in their asymptotic behavior, reflecting weaker interaction. As the mass increases, the deviation becomes more pronounced, indicating a stronger coupling effect. A similar pattern is observed when varying the adhesion frequency $f_{A}$ while keeping the resonator frequency $f_{r}$ fixed, as shown in Figure \ref{fig:figSI}(e). In this case, decreasing the thickness of the adhesive layer (which increases $f_{A}$) causes the two branches to move further away from their uncoupled frequencies, again due to enhanced coupling.

Furthermore, the spatial density of the resonators, represented by $A_{r}$, also significantly influences the coupling. This is because the coupling term in Equation \ref{eq:eq15a} ($q_r A_r/(\rho_l h)$) directly depends on $A_{r}$. Figure \ref{fig:figSI}(f) illustrates how $f_{r1}$ and $f_{r2}$ vary with changes in resonator spacing, which is inversely proportional to the density $A_r$. The figure compares two regimes: weak coupling ($f_{r}$ $<<$ $f_{A}$, red curves) and strong coupling ($f_{r}$ $=$ $f_{A}$, blue curves). In the strong coupling case, the repulsion between the branches is significantly more pronounced than in the weak case. This behavior clearly demonstrates that both the resonant frequencies and the spatial arrangement of the resonators play a vital role in determining the nature and intensity of the coupling with the adhesive layer.

\subsection{Finite Element modeling of bilayers for adhesion-induced resonant attenuation of SAWs}

The FE model used for the simulations performed in Section III of the main text is based on the model described in the Section I.B, with a few modifications. To reproduce transmission measurements that can be validated experimentally, we coated the entire width of the glass substrate with a 100 nm thick gold film. This coating acts as an optoacoustic transducer at the surface of the substrate, and is essential to perform optoacoustic measurements with techniques such as the laser ultrasonics \cite{sensing3}. On top of this gold layer, which mimics the optoacoustic transducer, an interfacial PMMA and an adhesive gold layer with variable thicknesses are placed at the center of the substrate over a width of 20 $\lambda_{min}$, $\lambda_{min}$ being the Rayleigh-wave wavelength in the glass substrate at $f = 1.5$ GHz.

As in Section I.B of the main text, a frequency response study is carried out. The substrate dimensions (width and depth) are scaled with the Rayleigh-wave  wavelength in the substrate over a frequency range from $f = 100$ MHz to $f = 1.5$ GHz with a 1 MHz step. The magnitude of the normal displacement after the interfacial and adhesive layers is obtained by placing a probe over a surface region with a width of $2 \lambda$ and a depth of $\lambda$. The transmission is computed as the ratio between the magnitude of the vertical displacement in the presence and in the absence of the interfacial and adhesive layers. The simulations are repeated for different thicknesses of the interfacial layer (PMMA) and of the adhesive layer (gold), as described in Section III of the main text.

The material parameters used in our simulations are the following: for gold, $c_{L,g} = 3200$~m/s, $c_{T,g} = 1200$~m/s, and $\rho_g = 19300$~kg/m$^3$; for the glass substrate, $c_{L,s} = 5700$~m/s, $c_{T,s} = 3400$~m/s, and $\rho_s = 2500$~kg/m$^3$ \cite{sensing3}; and for PMMA, $c_{L,\mathrm{p}} = 2740$~m/s, $c_{T,\mathrm{p}} = 1390$~m/s, and $\rho_{\mathrm{p}} = 1180$~kg/m$^3$ \cite{pmmafilm2,pmmafilm3}.

\nocite{*}

\end{document}